\documentclass[11pt]{article}

\usepackage{bibentry}          
\usepackage{attrib}            
\usepackage{bm}
\usepackage{setspace}
\usepackage{graphicx,scalerel}
\usepackage{amsmath}
\usepackage{array}
\usepackage{url}
\usepackage{amssymb}
\usepackage{caption}
\usepackage{float}
\usepackage{enumitem} 
\usepackage{subcaption}
\usepackage{natbib}
\usepackage{caption}
\usepackage{framed}
\usepackage{multirow} 
\usepackage{setspace}
\usepackage{color}
\usepackage{tikz}
\usepackage{pgfgantt}
\usepackage{mathtools}
\usepackage{nccmath}
\usepackage{longtable}
\usepackage[utf8]{inputenc} 
\usepackage[T1]{fontenc} 

    {\end{pmatrix}\end{medsize}}%
    
\newcommand\sbullet[1][.5]{\mathbin{\ThisStyle{\vcenter{\hbox{%
  \scalebox{#1}{$\SavedStyle\bullet$}}}}}%
}

\usepackage{listings}

\usepackage[]{graphicx}\usepackage[]{xcolor}
\makeatletter
\def\maxwidth{ %
  \ifdim\Gin@nat@width>\linewidth
    \linewidth
  \else
    \Gin@nat@width
  \fi
}
\usepackage{amsmath}
\usepackage{amssymb}
\usepackage{graphicx,scalerel}

\usepackage{fullpage}

\usepackage{orcidlink,thumbpdf,lmodern}

\usepackage{amsmath}
\usepackage{amssymb}
\usepackage{bm}
\usepackage{scalerel}
\usepackage{scalerel}
\usepackage{longtable}

\begin{document}

\begin{center}
{\large{\textbf{\texttt{fdesigns}: Bayesian Optimal Designs of Experiments for Functional Models in \texttt{R}}} } \\[1ex] 
		
D. Michaelides\footnote{Damianos Michaelides; dm3g15@soton.ac.uk; School of Mathematical Sciences, University of Southampton, Southampton, SO17 1BJ, UK.}
, A. M. Overstall \& D. C. Woods \\[1ex]
		
Southampton Statistical Sciences Research Institute \\
University of Southampton, United Kingdom
\end{center}			
				

\normalsize

This paper describes the \texttt{R} package \texttt{fdesigns} that implements a methodology for identifying Bayesian optimal experimental designs for models whose factor settings are functions, known as profile factors. This type of experiments which involve factors that vary dynamically over time, presenting unique challenges in both estimation and design due to the infinite-dimensional nature of functions. The package \texttt{fdesigns} implements a dimension reduction method leveraging basis functions of the B-spline basis system. The package \texttt{fdesigns} contains functions that effectively reduce the design problem to the optimisation of basis coefficients for functional linear functional generalised linear models, and it accommodates various options. Applications of the \texttt{fdesigns} package are demonstrated through a series of examples that showcase its capabilities in identifying optimal designs for functional linear and generalised linear models. The examples highlight how the package's functions can be used to efficiently design experiments involving both profile and scalar factors, including interactions and polynomial effects.

\begin{flushleft}
Keywords: profile factors, dynamic factors, design of experiments, optimal designs, optimal designs in \texttt{R}, functional designs, A-optimality, D-optimality, pseudo-Bayesian optimal designs, functional linear models, functional generalised linear models, functional logistic models, functional Poisson models, basis functions, B-splines
\end{flushleft}


\section{Introduction} \label{sec:intro}

Experimentation is a common process of gathering data, under controlled conditions, across numerous fields of applied science, including medicine, biology, chemistry, and engineering. Design of Experiments (DoE) is an efficient way in statistics to structure effective experiments, and as a result, leverage the most available knowledge about specific applications. This concept has been extensively explored in the statistical literature for over a century, with famous references being the books by \citet{atkinson:2007}, \citet{antony:2014}, \citet{dean:2017}, and \citet{montgomery:2017}.

Designs that maximise the amount of information obtained from an experiment are referred to as optimal experimental designs. Optimal designs are explained in detail by \citet{wu:2011}, \citet{montgomery:2017}, and \citet{dean:2017}. Identifying optimal experimental designs requires a clear understanding of the experiment's purpose, as this purpose must be reflected in the design itself. This is typically achieved through the application of design criteria, defined via objective functions. 

DoE fits naturally in the Bayesian framework since the choice of the design is a process taking place prior to experimentation, and hence, data collection. Within the Bayesian framework, the objective of the experiment is expressed through a utility function that defines the gain of the experimenter from using a design $\bm{\xi}$ to obtain responses $\bm{y}$, assuming values for the parameters $\bm{\theta}$. Consequently, a Bayesian optimal design that belongs to the design space $\mathcal{X}$, i.e., $\bm{\xi}^{*} \in \mathcal{X}$, is one that maximises the expected utility with respect to the joint distribution of the unknown responses and parameters, as described by \citet{chaloner:1995},
\begin{equation}
\label{eq:expectedutility}
\begin{split}
\Psi(\bm{\xi}) & = \mathbb{E}_{\bm{\theta},\bm{y}}[u(\bm{\theta}, \bm{y}, \bm{\xi})] \\
& = \int_{\bm{\theta}} \int_{\bm{y}} u(\bm{\theta}, \bm{y}, \bm{\xi}) \pi(\bm{y}|\bm{\theta},\bm{\xi})  \pi(\bm{\theta}) \; d\bm{y}  \; d\bm{\theta}. 
\end{split}
\end{equation}

The vast majority of the optimal design methodologies in the statistics literature focus on experimental designs which typically involve setting up a treatment with fixed or static conditions of the controllable factors for each experimental run, where in each run the factors are varied simultaneously. However, technological advancements have given rise to experiments for models that involve profile factors \citet{morris:2015}, also referred to as dynamic factors in literature. Profile factors are defined as factors that have the ability to vary as a function of an indexing variable, typically time, within a single experimental run. In the context of experiments that involve profile factors, the design of experiments problem is to choose the right function to vary the profile factor, or factors, in every run of the experiment.

A significant challenge faced is the generality of the function space. Unlike experiments with static factors, where the design space is relatively well-defined, profile factors introduce a broader and more complex set of possible functions, which in general are infinite dimensional objects. 

In literature, \citet{georgakis:2013} adapted a response surface methodology to address time-varying factors through the application of dimension reduction techniques. This approach was expanded by \citet{roche:2015}, \citet{klebanov:2016}, and \citet{roche:2018}. Their work adopts a data-driven methodology, typically employing a second-order model, to facilitate experimentation with profile factors. In the initial work by \citet{georgakis:2013}, shifted Legendre polynomial bases have been used. Later, \citet{roche:2015} and \citet{roche:2018} introduced the use of Fourier bases as well as data-driven bases derived from Principal Component Analysis (PCA) and Partial Least Squares (PLS), operating under the assumption that pilot data are available before the experiment begins.

In a more recent paper, \citet{michaelides2021optimal} presented a methodology that can be used to find optimal functions for profile factors, assuming a functional linear model. As will be described in Section \ref{sec:flm}, the statistical model they considered has the ability to handle multiple profile factors, where the parameters requiring estimation are themselves functions of time. Their approach leverages the power of the standard linear model optimal design methodology, and its flexibility allows designs to be obtained for various different scenarios and optimality criteria. In summary, with more details in Section \ref{subsec:basesfuncs}, the methodology developed in \citet{michaelides2021optimal} utilises the use of bases functions, with focus on B-splines, to restrict the function space of the profile factor and parameter functions and overcome issues in parameter estimation and design. 

This paper describes the \texttt{R} \citep{R} package \texttt{fdesigns} that stands for functional designs and implements the optimal design of experiments methodology for functional linear models as in \citet{michaelides2021optimal}, with details in Section \ref{sec:flm}, and expands to the more complicated and computationally expensive functional generalised linear models as in \citet[Chapter 7]{michaelides:2023}, with details in Section \ref{sec:fglm}. Although there are several \texttt{R} packages available that identify (optimal) designs for several types of models, for instance, 
\begin{itemize} 
    \item[-] \texttt{acebayes} (\citep{overstall:2020}; \citep{overstall:2020package}), \texttt{AlgDesign} \citep{wheeler:2019}, \texttt{OptimalDesign} \citep{harman:2016}, and \texttt{ICAOD} \citep{masoudi:2022} provide functions for identifying exact and approximate designs for linear and non-linear models using various design criteria and algorithms, including exchange and integer programming algorithms,
    \item[-] \texttt{rodd} \citep{guchenko:2016} provides functions for identifying optimal discriminating designs, known as T-optimal designs, 
    \item[-] \texttt{skpr} \citep{morgan:2021} provides functions that can be used for identifying optimal split-plot designs, and 
    \item[-] \texttt{FrF2} \citep{gromping:2014} provides functions with main focus on fractional factorial designs, but also includes capabilities for generating and evaluating optimal designs, 
\end{itemize}
to the best of our knowledge \texttt{fdesigns} serves as the first \texttt{R} package for identifying Bayesian optimal experimental designs for functional linear and generalised linear models. The package is available on the Comprehensive \texttt{R} Archive Network (CRAN) and on GitHub at \url{https://cran.r-project.org/web/packages/fdesigns/index.html} and \url{https://github.com/damianosmichaelides/fdesigns}, respectively.

The package includes two primary functions named \texttt{pflm} and \texttt{pfglm}, a support function named \texttt{P}, and functions for printing and plotting class objects.

The \texttt{pflm} function, which stands for "parallel functional linear models", is used to find optimal experimental designs for functional linear models using the coordinate exchange algorithm \citep{meyer:1995}. Similarly, the \texttt{pfglm} function, which stands for "parallel functional generalised linear models", is used to find optimal experimental designs for functional generalised linear models, utilising the coordinate exchange algorithm and approximation methods \citep{gotwalt:2009}. The term "parallel" indicates that these functions have the ability to repeat the optimisation process across multiple starting designs \citep{goos:2011}. The support function \texttt{P} which stands for "polynomials", is used to compute profile factor polynomials of a specified degree. 

A significant part of the \texttt{fdesigns} code is implemented in \texttt{C++} using the package \texttt{Rcpp} \citep{eddelbuettel:2011} and the package \texttt{RcppArmadillo} \citep{eddelbuettel:2014}. 

The functional linear model and the functional generalised linear model along with their implementations by \texttt{pflm} and \texttt{pfglm} in \texttt{fdesigns} are described in Sections \ref{sec:flm} and \ref{sec:fglm}. The next section, Section \ref{sec:Pfunc}, elaborates on the use of the support function \texttt{P} and distinguishes it from built-in \texttt{R} functions that compute polynomial factors. Several examples for functional linear and functional generalised linear models implemented by \texttt{pflm} and \texttt{pfglm} to illustrate their capabilities are demonstrated in Section \ref{sec:examples}.

\section[Functional linear model and pflm]{Functional linear model and \texttt{pflm}} \label{sec:flm}

In experiments for models that depend on profile factors, a time interval from time $0$ to time $\mathcal{T}$ is assumed. For an experiment consisting of $n$ runs, in the $i^{th}$ run of the experiment ($i=1,2,\dots,n$) the controllable functions of the profile factors are set, and scalar responses are recorded at time $\mathcal{T}$. To model the relationship between these scalar responses and the profile factors the functional linear model is used. 

The functional linear model considered in \texttt{fdesigns} is the simplified model from \citet{michaelides2021optimal} that omits some of the interaction terms by restricting the multivariate parameter functions using properties of the Dirac delta \citep{balakrishnan:2003}. The restricted format decreases the number of parameters to be estimated, and hence, decreases the computational burden. In context, for $J$ profile factors and at the $i^{th}$ run of the experiment, the form of the functional linear model considered is,
\begin{equation}
\label{eq:fullmodelfunctional}
y_{i} = \int_{0}^{\mathcal{T}} \bm{f}^{T}(\bm{x}_{i}(t)) \; \bm{\beta}(t)  \; dt + \epsilon_{i}, \quad i = 1,2, \dots, n.
\end{equation}
The $J \times 1$ vector $\bm{x}_{i}(t)$ represents the functions of the profile factors at the $i^{th}$ run of the experiment,
\begin{equation}
\label{eq:xit}
\begin{split}
\bm{x}_{i}^{T}(t) & = 
\begin{pmatrix}
x_{i1}(t) & x_{i2}(t) & \cdots & x_{iJ}(t)
\end{pmatrix}, \quad i = 1,2, \dots, n,
\end{split}
\end{equation} 
with each $x_{ij}(t), i=1,2,\dots,n, j=1,2,\dots,J$, the function of the $j^{th}$ profile factors at the $i^{th}$ run of the experiment. The structure of the linear predictor to specify the main effects, interactions, and polynomials takes place through the $Q \times 1$ function of the profile factors $\bm{f}^{T}(\bm{x}_{i}(t))$ with $Q$ representing the total number of terms in the model. Finally, the $Q \times 1$ vector $\bm{\beta}(t)$ represents the vector of the unknown functional parameters,
\begin{equation}
\label{eq:betait}
\bm{\beta}^{T}(t) = 
\begin{pmatrix}
\beta_{1}(t) & \beta_{2}(t) & \cdots & \beta_{J}(t)
\end{pmatrix},
\end{equation} 
with each functional parameter $\beta_{j}(t) : [0,\mathcal{T}] \rightarrow \mathbb{R}, j=1,\dots, J$ an unknown function of time $ 0 \leq t \leq \mathcal{T}$.

\subsection{DoE methodology via bases functions} \label{subsec:basesfuncs}

The model in \eqref{eq:fullmodelfunctional} involves estimating an infinite number of unknown parameters since functions, and consequently parameter functions, are infinite dimensional objects. However, the experiment provides only a finite number of observed responses, leading to an under-determined system with more unknowns than equations, resulting in infinite possible solutions that perfectly fit the data. To overcome this, the function space of the functional parameters is constrained using basis function expansions, as explained by \citet[p.~44]{ramsay:2005}. These expansions are represented as follows:
\begin{equation}
\label{eq:basesbnu}
\beta_{q}(t) = \sum_{l=1}^{n_{\beta,q}} \theta_{ql}b_{ql}(t) = \bm{b}_{q}^{T}(t) \bm{\theta}_{q}, \quad q=1,2, \dots, Q,
\end{equation}
where the functions $\bm{b}_{q}^{T}(t) = [b_{q1}(t), b_{q2}(t), \dots, b_{q n_{\beta,q}}(t)]$ are known bases functions and the vector $\bm{\theta}_{q}^{T} = (\theta_{q1}, \theta_{q2}, \dots, \theta_{qn_{\beta,q}})$ is a vector of unknown coefficients. As a result, estimating the unknown functional parameters simplifies to estimating $\sum_{q=1}^{Q} n_{\beta,q}$ coefficients. The case of scalar parameter or the intercept is represented by a single basis function which is constantly equal to one. 

To achieve optimal experimental conditions, the choice of the functions of the profile factors is critical. Depending on the experimental setup, the function space for each profile factor can either be general or restricted to specific classes of functions, such as polynomials of a certain degree or step functions with defined breakpoints. To ensure meaningful inference, it is necessary to impose some restrictions on the function space of the profile factors as well. These restrictions can be effectively implemented using basis function expansions as follows:
\begin{equation}
\label{eq:basesx}
x_{ij}(t) = \sum_{l=1}^{n_{x,j}} \gamma_{ijl}c_{jl}(t), \quad i=1,2, \dots, n,  j=1,2, \dots, J.
\end{equation}
A scalar factor is represented by a single basis function which is constantly equal to one, and consequently $x_{ij}$ becomes a scalar value that must be specified in each experimental run. The bases expansion for each profile factor can be expressed in vector form as, $\bm{x}_{\sbullet j}(t) = \bm{\Gamma}_{j} \bm{c}_{j}(t)$, with $\bm{x}_{\sbullet j}$ the function of the $j^{th}$ profile factor in every run of the experiment, known bases functions $ \bm{c}_{j}(t)$, and $\bm{\Gamma}_{j}$ a $n\times n_{x,j}$ coefficient matrix. 

By combining the bases expansions and the matrix form of the functional linear model, an extended form of a linear model is formed,
\begin{equation}
\label{eq:extendedmodel}
\begin{split}
\bm{y} & = \int_{0}^{\mathcal{T}} \bm{f}^{T}(\bm{X}(t)) \; \bm{\beta}(t)  \; dt + \bm{\epsilon} \\
& = \int_{0}^{\mathcal{T}} \bm{f}^{T}(\bm{X}(t)) \; \bm{b}^{T}(t)  \; dt \; \bm{\theta} + \bm{\epsilon} \\
& = \bm{Z\theta} + \bm{\epsilon},
\end{split}
\end{equation}
where, $\bm{y}$ is the $n \times 1$ vector of responses, $\bm{Z}$ is the $n \times \sum_{q=1}^{Q} n_{\beta,q}$ model matrix, $\bm{\theta}$ is the $\sum_{q=1}^{Q} n_{\beta,q} \times 1$ vector of unknown parameters, and $\bm{\epsilon}$ is the $n \times 1$ vector of independent error terms with mean zero and variance-covariance $\sigma^{2} \bm{I}_{n}$. The matrix $\bm{X}(t)$ is a $n \times J$ matrix that carries the functions of the profile factors for each run. The function $\bm{f}^{T}(\bm{X}(t))$ is a functional of the functions of the profile factors that needs to be specified to define the structure of the model, i.e., main effects, interactions, and polynomial effects.

The model matrix $\bm{Z}$ is partitioned into $Q$ blocks, with each block $\bm{Z}_{\sbullet q}$ ($q=1,2,\dots,Q$) corresponding to specific effects, such as main, quadratic, or interaction effects of the profile factors. For main effects, $\bm{Z}_{\sbullet q}$ is defined as an integral of the product of the profile factor and parameter bases functions. Quadratic and interaction effects are derived from an integral of the Kronecker product of the bases functions of the profile factors involved in the effect, multiplied by the effect's parameter bases functions. A detailed expansion of these integrals is given in \citet{michaelides2021optimal}. For B-spline bases for the functions of the profile factors and power or B-spline bases for the parameter functions, the integrals defining the model matrix are available in closed form solutions \citep[Chapter 5]{michaelides:2023}.

\subsection{Bayesian approach using a roughness penalty} \label{subsec:bayes}

The aim of an experiment, as briefly described in Section \ref{sec:intro} is encapsulated via a utility function. The utility function shows the experimenter's gain from using a design, for the functional linear model that is $\bm{\Gamma}$. A Bayesian optimal design $\bm{\Gamma}^{*}$ is the design that maximises the expected utility over the joint distribution of unknown responses and parameters, as in \eqref{eq:expectedutility} and \citet{chaloner:1995}. 

Most utility functions depend on posterior quantities. For the functional linear model, the likelihood function follows a normal distribution, and the prior assigned to the unknown parameters $\bm{\theta}$ and $\sigma^{2}$ is the conjugate normal-inverse-gamma (NIG) distribution,
\begin{align}
\label{eq:likelihoodandprior}
\pi(\bm{y}|\bm{\theta},\sigma^{2}) \sim N(\bm{Z\theta}, \sigma^{2}\bm{I}_{n}) \nonumber \\ 
\pi(\bm{\theta},\sigma^{2}) \sim NIG(\bm{\mu}, \bm{V}, a/2, b/2),
\end{align}
where $\bm{\mu}$ is the $\sum_{q=1}^{Q} n_{\beta,q}$ prior mean vector of $\bm{\theta}$, $\bm{V}$ is a known and symmetric $\sum_{q=1}^{Q} n_{\beta,q} \times \sum_{q=1}^{Q} n_{\beta,q}$ matrix, and $a$ and $b$ are hyperparameters. The posterior density of the model parameters is also a NIG distribution,\begin{equation}
\label{eq:posteriorflm}
\pi(\bm{\theta},\sigma^{2}|\bm{y}) \sim NIG(\bm{\theta}_{N}, \bm{V}_{N},a^{*}/2,b^{*}/2)
\end{equation}
with updated hyperparameters,
\begin{align}
\label{eq:hyperparametersflm}
\bm{V}_{N} & = (\bm{Z}^{T}\bm{Z} + \bm{V}^{-1})^{-1}   \nonumber \\
\bm{\theta}_{N} & =  \bm{V}_{N} (\bm{V}^{-1}\bm{\mu} + \bm{Z}^{T}\bm{y})  \nonumber \\
 a^{*} & = a + n    \nonumber \\
 b^{*} & = b + (\bm{\mu}^{T}\bm{V}^{-1}\bm{\mu} + \bm{y}^{T}\bm{y} - \bm{\theta}_{N}^{T}  \bm{V}_{N}^{-1} \bm{\theta}_{N}).
\end{align}

Two common utility functions, that are implemented in the \texttt{fdesigns} package in \texttt{R} are: 

\begin{enumerate}

\item Negative Squared Error Loss (NSEL)

NSEL is a utility function in quadratic form, that measures the expected discrepancy between the true and estimated parameters,
\begin{equation}
\label{eq:nsel}
u(\bm{\theta},\bm{y},\bm{\Gamma}) = - \big[\bm{\theta} - \mathbb{E}(\bm{\theta}|\bm{y},\bm{\Gamma})\big]^{T} \big[\bm{\theta} - \mathbb{E}(\bm{\theta}|\bm{y},\bm{\Gamma})\big],
\end{equation}
leading to an objective function that maximises the negative trace of the variance-covariance matrix of the posterior distribution (known as the Bayesian A-optimality),
\begin{equation}
\label{eq:nseloptimal}
\Psi_{NSEL}(\bm{\Gamma}) = - \text{tr} \big[ \big(\bm{Z}^{T}\bm{Z} + \bm{V}^{-1} \big)^{-1}\big].
\end{equation}

\item Shannon Information Gain (SIG)

SIG measures the information gain between the posterior and prior distributions,
\begin{equation}
\label{eq:sig}
\begin{split}
u(\bm{\beta},\bm{y},\bm{X}) & = \log \pi(\bm{\beta}|\bm{y}) - \log \pi(\bm{\beta}) \\
& = \log \pi(\bm{y}|\bm{\beta}) - \log \pi(\bm{y}),
\end{split}
\end{equation}
leading to an objective function that maximises the $\log$ determinant of the posterior variance-covariance matrix (known as Bayesian D-optimality),
\begin{equation}
\label{eq:sigoptimal}
\Psi_{SIG}(\bm{\Gamma}) = \log |\bm{Z}^{T}\bm{Z} + \bm{V}^{-1}|.
\end{equation}

\end{enumerate}

The selection of bases functions poses challenges, such as choosing the degree and number and placement of knots in splines. An alternative approach involves the use of a roughness penalty, which penalises the complexity of the parameter functions, thereby regularising the model with a smoothing 
parameter $\lambda$. This approach balances the trade-off between fit and smoothness, simplifying the problem to selecting a single smoothing 
parameter. The roughness penalty is typically defined in the quadratic form $\bm{\theta}^{T} \bm{R}_{0} \bm{\theta}$, where $\bm{R}_{0}$ is 
a $\sum_{q=1}^{Q} n_{\beta,q} \times \sum_{q=1}^{Q} n_{\beta,q}$ differentiating block diagonal matrix controlling the roughness of the parameter functions \citep{ramsay:2005}. 

The roughness penalty approach influences the formulation of the prior precision matrix which can be expressed as $\bm{V}^{-1} = \lambda \bm{R}_{0}$. In essence, $\lambda$ directly controls the prior specification where a value close to zero indicates a weak prior and minimal smoothing, while a large $\lambda$ imposes strong priors and heavy smoothing of the parameter functions. The choice of $\lambda=0$ is the special case of frequentist designs. 

The NSEL and SIG objective functions from equations \eqref{eq:nseloptimal} and \eqref{eq:sigoptimal} incorporate the smoothing parameter as follows,
\begin{equation}
\label{eq:nseloptimallambda}
\Psi_{NSEL}(\bm{\Gamma}) = \text{tr} \big(\bm{Z}^{T}\bm{Z} + \lambda \bm{R}_{0} \big)^{-1}
\end{equation}
\vspace{-0.5cm}
\begin{equation}
\label{eq:sigoptimallambda}
\Psi_{SIG}(\bm{\Gamma}) = \log |\bm{Z}^{T}\bm{Z} + \lambda \bm{R}_{0}|.
\end{equation}

\subsection[Implementation of pflm()]{Implementation of \texttt{pflm}} \label{subsec:pflm}

Functional linear models are implemented using the \texttt{pflm} function, which consists of several mandatory and optional arguments. The mandatory arguments are used to define the model structure and to specify the experimental settings, including the profile factors and functional parameters. In contrast, the optional arguments provide flexibility for the user to incorporate a smoothing parameter or adjust default bounds for time and profile factors. The function \texttt{pflm} is utilised as follows:
\begin{verbatim}
pflm(formula, nsd = 1, mc.cores = 1, npf, tbounds, nruns, 
      startd = NULL, dx, knotsx, pars, db, knotsb = NULL, 
      lambda = 0, criterion = c("A", "D"), dlbound = -1, 
      dubound = 1, tol = 0.0001, progress = FALSE)
\end{verbatim}
The arguments of the function \texttt{pflm} are described in Table~\ref{table:argumentspflm}. The use of asterisks in Table~\ref{table:argumentspflm} highlights the mandatory arguments. Polynomial effects in the \texttt{formula} argument need to be specified via the \texttt{fdesigns} supporting function \texttt{P} that is discussed in Section \ref{sec:Pfunc}.

\begin{longtable}{p{2.5cm}p{12cm}}
\hline
 \textbf{Argument} & \textbf{Description}  \\
 \hline
  \texttt{formula*} & Object of type formula, to create the model equation. Elements need to match the list names for \texttt{startd}. Main effects are called using the names of the profile factors in \texttt{startd}, interactions are called using the names of the profile factors in \texttt{startd} separated with :, and polynomial effects are called using the function \texttt{P}. Scalar factors are called using the same way and degree and knots through the arguments \texttt{dx} and \texttt{knotsx} are used to specify the scalar factors. A scalar factor is equivalent to a profile factor with degree 0 and no interior knots. \\
   \hline
\texttt{nsd} & The number of starting designs. The default entry is 1. \\
 \hline
\texttt{mc.cores} & The number of cores to use. The option is initialized from environment variable MC\_CORES if set. Must be at least one, and for parallel computing at least two cores are required. The default entry is 1. \\
 \hline
 \texttt{npf*} & The total number of (profile) factors in the model. \\
 \hline
 \texttt{tbounds*} & A time vector of length 2, representing the boundaries of time, i.e., 0 and T. \\
 \hline
 \texttt{nruns*} & The number of runs of the experiment. \\
 \hline
 \texttt{startd} & Representing the starting design but if NULL then random designs are automatically generated. It should be a list of length \texttt{nsd}, and each component should be a list of length \texttt{npf}. \\
 \hline
 \texttt{dx*} & A vector of length \texttt{npf}, representing the degree of B-spline basis functions for the functions of the functional factors. A scalar factor must have a zero degree entry. \\
 \hline
 \texttt{knotsx*} & A list of length \texttt{npf}, with every object in the list representing the knot vectors of each functional factor. A Scalar factor must have no interior knots, i.e., an empty knot vector. \\
 \hline
 \texttt{pars*} & A vector of length equal to the total terms in formula, representing the basis choice for the (functional) parameters. Entries should be "power" or "bspline". A scalar parameter is represented through a "power" basis. \\
 \hline
 \texttt{db*} & A vector of length equal to the total terms in formula, representing the degree of the basis for
the (functional) parameters. For power series basis the degree is: 1 for linear, 2 for quadratic, etc. A scalar parameter must have degree 0. \\
 \hline
 \texttt{knotsb} & A list of length equal to the total terms in formula, representing the knot vector of each (functional) parameter. For parameters represented by a power series basis, the knot vector should be empty or NULL. \\
 \hline
 \texttt{lambda} & Smoothing parameter to penalise the complexity of the functions of the profile factors. The default value is 0, i.e., no penalty. \\
 \hline
 \texttt{criterion*} & The choice of objective function. Currently there are two available choices: A-optimality (\texttt{criterion} = "A") and D-optimality (\texttt{criterion} = "D"). \\
 \hline
 \texttt{tol} & The tolerance value in the optimisation algorithm. Default value is 0.0001. \\
 \hline
 \texttt{dlbound} & The design's lower bound. The default lower bound is -1. \\
 \hline
 \texttt{dubound} & The design's upper bound. The default upper bound is 1. \\
 \hline
 \texttt{progress} & If TRUE, it returns the progress of iterations from the optimisation process. The default entry is FALSE. \\
 \hline
\caption{The arguments of the function \texttt{pflm} with a description for each. The asterisk * indicates a mandatory argument.}
\label{table:argumentspflm}
\end{longtable}

The \texttt{pflm} function returns an object of class \texttt{flm} which is a list comprising multiple components. The most important output component is the final design (\texttt{design}) and its corresponding final objective value (\texttt{objval}). Additional output components include the number of iterations required to obtain the final design (\texttt{nits}), the initial design that led to the final outcome (\texttt{startd}), and several of 
the input arguments. When the procedure involves multiple starting designs, the output also includes the index of the repetition that resulted in 
the final design (\texttt{bestrep}), as well as lists of all initial designs (\texttt{allstartd}), all final designs (\texttt{alldesigns}), and all final objective values (\texttt{allobjvals}). A comprehensive list of the output components of the \texttt{pflm} function, along with brief descriptions of each, is provided in Table~\ref{table:outputspflm} in Appendix \ref{app:tables}.

Printing the resulting \texttt{flm} object provides a summary of the inputs, the objective value, the number of iterations, and the computing resources used, such that:

\begin{verbatim}
R> print()
\end{verbatim}

or

\begin{verbatim}
R> summary()
\end{verbatim}

\begin{verbatim}
The number of profile factors is: ()
The number of runs is: ()
The objective criterion is: ()
The objective value is: ()
The number of iterations is: ()
The computing elapsed time is: ()
\end{verbatim}

Examples of \texttt{pflm} are demonstrated in Section \ref{sec:examples}.

\section[Functional generalised linear model and pfglm]{Functional generalised linear model and \texttt{pfglm}} \label{sec:fglm}

Functional generalised linear models represent the relationship between responses that belong to the exponential family of distributions \citep[p. 103]{wood:2017} and functions of profile factors \citep{marx:1999}, 
\begin{flalign}
\label{eq:fglmfullmodelmatrix}
& \bm{y} \sim \text{EFD}(\bm{\mu}, \phi \bm{A}) \nonumber \\
& g(\bm{\mu}) = \int_{0}^{\mathcal{T}} \bm{f}^{T}(\bm{X}(t)) \; \bm{\beta}(t)  \; dt = \bm{\eta},
\end{flalign}
with $\bm{f}^{T}(\bm{X}(t))$ and $\bm{\beta}(t)$ as defined in model \eqref{eq:extendedmodel}, $\bm{\mu}$ the $n \times 1$ vector containing the mean of the responses, $\phi$ the dispersion parameter, and $\bm{A}$ the $n \times 1$ vector of the weights of the dispersion parameter with the $i^{th}$ entry being $1/w_{i},  i = 1,2, \dots, n$. 

Since the linear predictor of the generalised model in \eqref{eq:fglmfullmodelmatrix} is identical to the linear predictor \eqref{eq:extendedmodel}, the same methodology with bases function expansions to restrict the function space is followed. In that case, the linear predictor of the generalised model updates to,
\begin{equation}
\label{eq:extendedmodelglm}
g(\bm{\mu}) = \bm{Z\theta} 
\end{equation}
with $\bm{Z}$ the $n \times \sum_{q=1}^{Q} n_{\beta,q}$ model matrix that is partitioned in $Q$ column blocks and $\bm{\theta}$ the $\sum_{q=1}^{Q} n_{\beta,q} \times 1$ vector of unknown parameters from \eqref{eq:extendedmodel}.

\subsection{Pseudo Bayesian designs} \label{subsec:pseudo}

As discussed in the previous section, the response follows an exponential family distribution. Given that, an asymptotic approximation to the posterior variance-covariance matrix is given by the inverse of the Fisher information matrix \citep{chaloner:1995}. By maximising the penalised log-likelihood, which is the log-likelihood function penalised by the roughness penalty $\lambda \bm{\theta}^{T} \bm{R}_{0} \bm{\theta}$, the Fisher information matrix is,
\begin{equation}
\label{eq:infomatrixfglmroughness}
\mathcal{I}(\bm{\theta}, \bm{\Gamma}) = \bm{Z}^{T} \bm{W} \bm{Z} + \lambda \bm{R}_{0}
\end{equation}
where $\bm{W}$ is a $n\times n$ diagonal matrix with entries,
\begin{equation*}
w_{ii} = \frac{1}{g'(\mu_{i})^{2}} \; \frac{1}{\text{Var}(y_{i})}, \quad i=1,2, \dots, n.
\end{equation*}
Identifying optimal designs is challenging because the Fisher information matrix, and consequently the design itself, depends on the unknown parameters $\bm{\theta}$. For this reason, optimal experimental designs for functional generalised linear models require prior information about the model parameters. To incorporate prior information and define the optimality objective functions, the pseudo-Bayesian approach is adopted (\citep{chaloner:1995}; \citep{overstall:2017}). For instance, approximations to the NSEL and SIG utility functions are,
\begin{equation}
\label{eq:nselapprx}
\tilde{\Psi}_{NSEL}(\bm{\Gamma}) = - \text{tr} \big(\mathcal{I}(\bm{\theta}, \bm{\Gamma})^{-1} \big),
\end{equation}
\vspace{-0.5cm}
\begin{equation}
\label{eq:sigapprx}
\tilde{\Psi}_{SIG}(\bm{\Gamma}) = \log |\mathcal{I}(\bm{\theta}, \bm{\Gamma})|.
\end{equation}
known as pseudo-Bayesian A-optimality and pseudo-Bayesian D-optimality, and designs that maximise the expectation of the pseudo-Bayesian A- and D- optimality objective functions are known as pseudo-Bayesian A- and D-optimal designs, respectively \citep{van:2014}.

However, the expectation of the objective functions with respect to the prior distribution results from analytically intractable, and usually high-dimensional integrals that need to be approximated numerically. The methods implemented in \texttt{fesigns} are the Monte Carlo (MC) and quadrature approximations. 

The Monte Carlo approximation is a stochastic method that evaluates numerically the expectation of a general function, and relies on generating random samples from the prior distribution \citep{caflisch:1998}, 
\begin{equation}
\label{eq:MCapproximation}
\int_{\bm{\Theta}} \psi(\bm{\theta}, \bm{\Gamma}) \: \pi(\bm{\theta}) \; d\bm{\theta} \approx \frac{1}{B} \sum_{b=1}^{B} \psi(\bm{\theta_{b}}, \bm{\Gamma}),
\end{equation}
with $\psi(\bm{\theta_{b}}, \bm{\Gamma})$ the function evaluated at the parameter random sample $\bm{\theta_{b}}$ for $b=1,2,\dots,B$. As the number of random samples becomes larger, the approximation gets closer to the actual expectation \citep{lapeyre:2007},
\begin{equation*}
\frac{1}{B} \sum_{b=1}^{B} \psi(\bm{\theta_{b}}, \bm{\Gamma}) \rightarrow \mathbb{E}_{\bm{\theta}} \Big\{ \psi(\bm{\theta}, \bm{\Gamma}) \Big\}, \quad \text{as} \; \: B \rightarrow \infty.
\end{equation*}
In \texttt{fesigns}, the size of Monte Carlo samples is specified via the argument \texttt{B} that defaults to $10000$ samples. 

The quadrature approximation is a deterministic method that is a weighted version of the MC approximation, with the expectation evaluated at selected points,
\begin{equation}
\label{eq:quadratureapproximation}
\int_{\bm{\Theta}} \psi(\bm{\theta}, \bm{\Gamma}) \: \pi(\bm{\theta}) \; d\bm{\theta} \approx \sum_{b=1}^{B} \omega_{b} \: \psi(\bm{\theta_{b}}, \bm{\Gamma}),
\end{equation}
with $\bm{\theta_{b}}$ and $\omega_{b}$, for $b=1,2,\dots,B$, being abscissas and weights (\citep{gotwalt:2009}; \citep[Chapter 10]{gelman:2013}). 

In \texttt{fesigns}, prior distributions implemented are the normal and uniform distributions. For normal priors, the Gauss-Hermite quadrature rule is followed \citep{salzer:1952} and for uniform priors, the Gauss-Legendre quadrature rule is followed (\citep{lether:1978}; \citep{hale:2013}). Both quadrature rules used in \texttt{fesigns} are formulated in the \texttt{mvQuad} package \citep{mvQuad}. High dimensional integrals require a large number of quadrature points for accurately approximation, hence, increasing the computational cost.

\subsection[Implementation of pfglm()]{Implementation of \texttt{pfglm}} \label{subsec:pfglm}

Functional generalised linear models are implemented using the \texttt{pfglm} function. Most of the arguments for \texttt{pfglm} are the same with arguments of the \texttt{pflm} function. This is because both \texttt{pflm} and \texttt{pfglm} utilise the same approach for specifying the model structure, experimental settings, profile factors, and functional parameters. On top of the same arguments, \texttt{pfglm} includes additional mandatory and optional arguments that are used for defining the response family, incorporating prior information, and specifying the choice of the numerical approximation method for evaluating the expectation of the objective function with respect to the prior distribution. The function \texttt{pfglm} is utilised as follows:
\begin{verbatim}
pfglm(formula, nsd = 1, mc.cores = 1, npf, tbounds, nruns,
        startd = NULL, dx, knotsx, pars, db, knotsb = NULL,
        lambda = 0, criterion = c("A", "D"), family,
        method = c("quadrature", "MC"), level = NULL, B = NULL,
        prior, dlbound = -1, dubound = 1, tol = 0.0001,
        progress = FALSE)
\end{verbatim}
The additional arguments of the function \texttt{pfglm}, that are not described in Table~\ref{table:argumentspflm} for \texttt{pfglm}, are described in Table~\ref{table:argumentspfglm}. The use of asterisks in Table~\ref{table:argumentspfglm} highlights the mandatory arguments. Polynomial effects in the \texttt{formula} argument need to be specified via the \texttt{fdesigns} supporting function \texttt{P} that is discussed in Section \ref{sec:Pfunc}.

\begin{longtable}{p{2.5cm}p{12cm}}
\hline
 \textbf{Argument} & \textbf{Description}  \\
 \hline
  \texttt{family*} & Specifies the error distribution and the link function of the functional generalised linear model. It can be the name of a 
  family in the form of a character string, or an \texttt{R} \texttt{family} function. Currently, the methodology is implemented only for the 
  binomial family with the logit link, i.e., \texttt{family} = binomial(link = "logit"), and the Poisson family with the log link, 
  i.e., \texttt{family} = poisson(link = "log"). \\
 \hline
  \texttt{method*} & A character argument specifying the method of approximation of the expectation of the objective function with respect to a prior distribution of the parameters. Currently there are two available choices: 1. Deterministic quadrature approximation (\texttt{method} = "quadrature"); and 2. Stochastic Monte Carlo approximation (\texttt{method} = "MC"). \\
\hline
 \texttt{level} & An optional argument that specifies the accuracy level in the quadrature approximation. It is the number of points in each dimension. If NULL and \texttt{method} = "quadrature", then it defaults to 5. A high value of level may increase the computation time; especially for complicated models. If the model is complicated, i.e., several profile factors or interactions and polynomials, prefer to use \texttt{method} = "MC". \\
 \hline
  \texttt{B} & An optional argument that specifies the size of the Monte Carlo samples. If NULL and \texttt{method} = "MC", then it defaults to 10000. For \texttt{method} = "quadrature", B is computer automatically according to the dimensionality of the functional model and the \texttt{level} argument. \\
 \hline
 \texttt{prior*} & An argument to specify the prior distribution. For \texttt{method} = "MC", it should be a function of two arguments \texttt{B} and \texttt{Q}. Both arguments are integers. The value of \texttt{B} corresponds to the argument \texttt{B}, and the value of \texttt{Q} represents the total number of basis functions of the functional parameters. The function must generate a matrix of dimensions \texttt{B} by \texttt{Q}, that contains a random sample from the prior distribution of the parameters. For \texttt{method} = "quadrature", normal and uniform prior distribution for the parameters are allowed. For a normal prior distribution, the argument prior needs to be a list of length 2, with the entries named "mu" for the prior mean and "sigma2" for the prior variance-covariance matrix. The prior mean can be a scalar value that means all parameters have the same prior mean, or a vector of prior means with length equal to the number of parameters in the functional model. The prior variance-covariance can be a scalar value that means all parameters have a common variance, or a vector of prior variances with length equal to the number of parameters in the functional model, or a square matrix with the number of rows and columns equal to the number of parameters in the functional model. For a uniform prior distribution, the argument prior needs to be a list of a single entry named "unifbound" for the lower and upper bounds of the prior distribution. The bounds can be a vector of length 2 that means all parameters have the same bounds, or a matrix with the number of rows equal to 2 and the number of columns equal to the number of parameters in the functional model. \\
 \hline
\caption{The additional arguments of the function \texttt{pfglm}, that are not used in the function \texttt{pflm}, with a description for each. 
The asterisk * indicates a mandatory argument.}
\label{table:argumentspfglm}
\end{longtable}

The \texttt{pfglm} function returns an object of class \texttt{fglm} which is a list comprising multiple components. Most of the components are identical to the output of \texttt{pflm}, but with additional components, which are the family, the prior specification, and the method of approximation. A comprehensive list of the output components of the \texttt{pfglm} function, along with brief descriptions of each, is provided in Table~\ref{table:outputspfglm} in 
Appendix \ref{app:tables}.

Printing the resulting \texttt{fglm} object provides a summary of the inputs, the objective value, the number of iterations, and the computing 
resources used, such that:

\begin{verbatim}
R> print()
\end{verbatim}

or

\begin{verbatim}
R> summary()
\end{verbatim}

\begin{verbatim}
The number of profile factors is: ()
The number of runs is: ()
The objective criterion is: ()
The objective value is: ()
The number of iterations is: ()
The method of approximation is: ()
The family distribution and the link function are: () and ()
The computing elapsed time is: ()
\end{verbatim}

Examples of \texttt{pfglm} are demonstrated in Section \ref{sec:examples}.

\section[Implementation of P for polynomial effects]{Implementation of \texttt{P} for polynomial effects} \label{sec:Pfunc}

The function \texttt{P} in \texttt{fdesigns} is specifically designed for handling profile factor polynomial effects. It extends the capabilities of 
the base function \texttt{I} to accommodate profile factors, which differ from static factors by being expressed as linear combinations of 
basis functions. Hence, \texttt{P} is needed for the calculation of higher-order polynomials of profile factors through the product of 
the basis function combinations corresponding to the profile factor involved in the polynomial term.

As a support function within the package, \texttt{P} is essential in defining polynomial effects within the mandatory \texttt{formula} argument of the primary functions \texttt{pflm} and \texttt{pfglm}. Its usage is defined as follows:

\begin{verbatim}
P(x, deg)
\end{verbatim}

with both of its arguments being mandatory. The arguments of the function \texttt{P} are described in Table~\ref{table:argumentsP}. The use of 
asterisks in Table~\ref{table:argumentsP} highlights the mandatory arguments.

\begin{longtable}{p{2.5cm}p{12cm}}
\hline
 \textbf{Argument} & \textbf{Description}  \\
 \hline
  \texttt{x*} & A coefficient matrix from the basis expansion of a profile factor. The name passed needs to match the name of the profile factor 
  from the argument \texttt{startd} in \texttt{pflm} and \texttt{pfglm}. \\
   \hline
\texttt{deg*} & The degree of the polynomial effect for the profile factor. \\
 \hline
\caption{The arguments of the function \texttt{P} with a description for each. The asterisk * indicates a mandatory argument.}
\label{table:argumentsP}
\end{longtable}

The output of \texttt{P} is an attributes list that contains the polynomial coefficient matrix of the profile factor involved in the polynomial 
term, the argument \texttt{x}, and the argument \texttt{deg}.

\section[Application of fdesigns]{Application of \texttt{fdesigns}} \label{sec:examples}

This section serves to illustrate the application of \texttt{fdesigns} through several examples, focusing primarily on the core functions, \texttt{pflm} and \texttt{pfglm}. The support function \texttt{P} is utilised within the \texttt{formula} argument of these main functions to specify polynomial effects in the functional models and compute the profile factor polynomials. The examples cover a range of scenarios involving functional linear and functional generalised linear models.

While \texttt{pflm} and \texttt{pfglm} are computationally efficient, identifying optimal designs for complex models that depend on multiple profile 
factors can increase the computational demand, and hence, necessitating greater computational resources. For this reason, the examples presented are intentionally kept relatively simple to illustrate \texttt{fdesigns}'s functionality effectively.

\subsection{Functional linear regression model} \label{subsec:example1}

An experiment is assumed to be modeled by a functional linear model that depends on a single profile factor. The aim is to identify a Bayesian D-optimal design of $n=4$ runs, thus, to appropriately choose the functions of the profile factor in each run of the experiment. Given a single profile factor is of interest, the functional linear model is defined as:
\begin{equation}
\label{eq:exampleA}
y_{i} = \beta_{1} + \int_{0}^{\mathcal{T}} \beta_{2}(t) x_{i1}(t) \; dt + \epsilon_{i}, \quad i = 1,2,3,4, \;  t \in [0,1].
\end{equation}

The experimental capabilities are rump functions, so control of the profile factor is achieved by a linear B-spline, i.e., degree one B-spline, with two equally spaced interior knots. The number of runs, number of factors and specifications of the factor will be passed to \texttt{pflm} using:

\begin{verbatim}
R> nruns <- 4
R> npf <- 1
R> dx <- c(1)
R> knotsx <- list(c(0.333, 0.666))                				
\end{verbatim}

For the functional parameter that corresponds to the profile factor, a quadratic power series basis is used. A prior specification is set via the roughness penalty approach to penalise the complexity of the functions with a smoothing parameter $\lambda=10$. The above information will be passed to \texttt{pflm} using:

\begin{verbatim}
R> pars <- c("power")
R> db <- c(2)
R> knotsb <- list(c())   
R> lambda <- 10
\end{verbatim}

The Bayesian D-optimal design search will be for \texttt{nsd}$=100$ random starts that will be generated automatically in \texttt{pflm} by leaving \texttt{startd} in its default option. For reproducibility a \texttt{set.seed} is used. All other arguments remain at their default options. The Bayesian D-optimal design is identified via the \texttt{pflm} such that:

\begin{verbatim}
R> set.seed(0)
R> exampleAa <- pflm(formula = ~ x1, nsd = 100, npf = npf,
+                tbounds = c(0, 1), nruns = nruns, dx = dx, 
+                knotsx = knotsx, pars=pars, db = db, 
+                knotsb = knotsb, criterion = "D", lambda = lambda)
\end{verbatim}

and the main details about the resulting \texttt{flm} object are printed via:

\begin{verbatim}
R> print(exampleAa)
\end{verbatim}

\begin{verbatim}
The number of profile factors is: 1

The number of runs is: 4

The objective criterion is: D-optimality

The objective value is: 0.4051947

The number of iterations is: 4

The computing elapsed time is: 00:00:00
\end{verbatim}

\begin{figure}[ht]
\centering
\includegraphics[width=\linewidth]{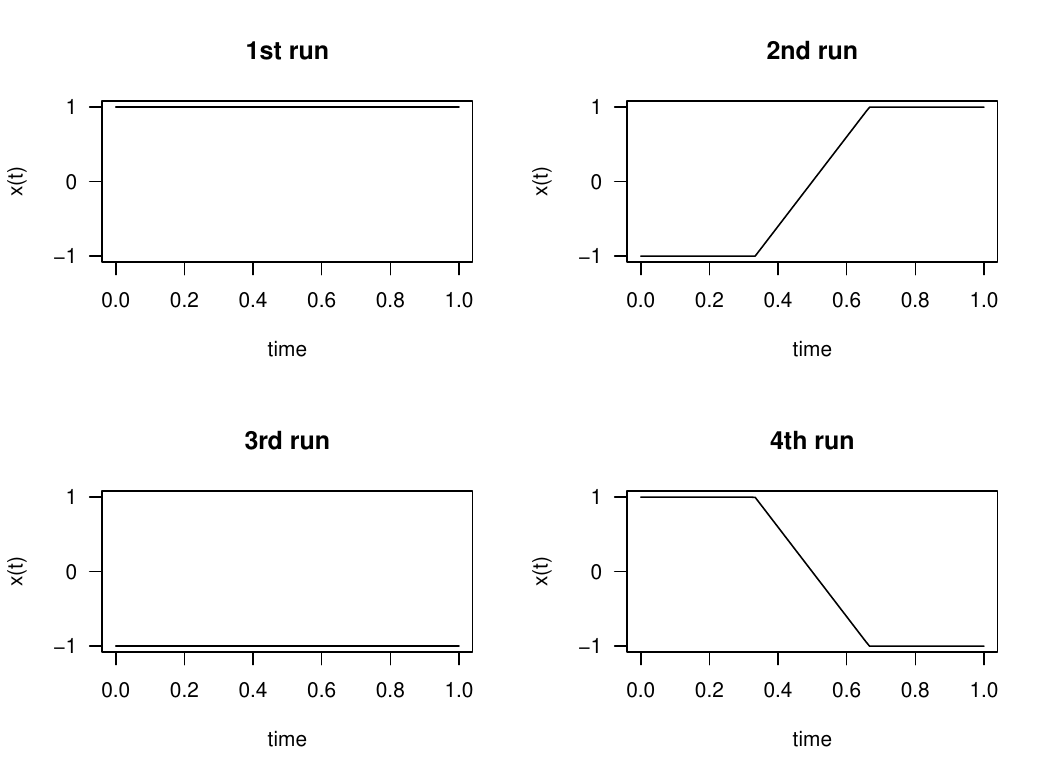}
\caption{Bayesian four-run D-optimal design for the functional linear regression model and quadratic prameter basis.}
\label{fig:exampleAa}
\end{figure}

The Bayesian D-optimal design identified can be accessed via calling the design component of the \texttt{flm} object such that:

\begin{verbatim}
R> exampleAa$design
\end{verbatim}

and the functions of the profile factor resulting from the Bayesian four-run D-optimal design identified are plotted via:

\begin{verbatim}
R> plot(exampleAa)
\end{verbatim}

and then input the number of the profile factor to plot (in this case 1) to the question appearing "Which profile factor to plot?". The functions of the profile factor as shown in Figure~\ref{fig:exampleAa}.

\begin{figure}[ht]
\centering
\includegraphics[width=\linewidth]{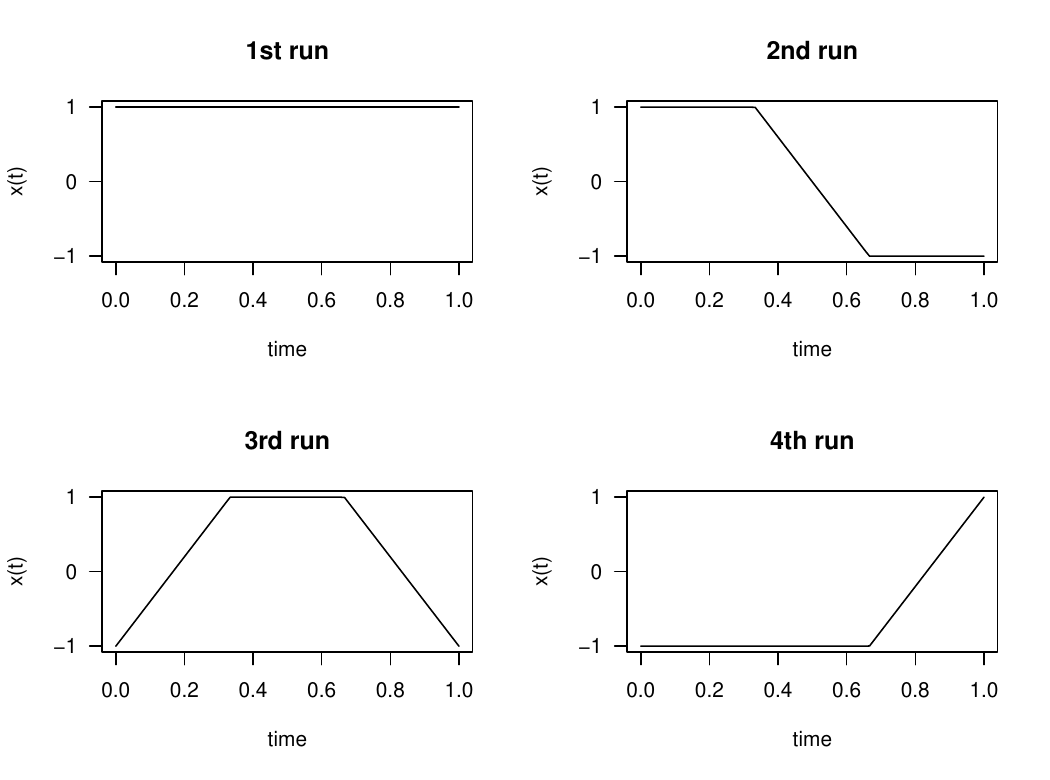}
\caption{Bayesian four-run D-optimal design for the functional linear regression model and linear B-spline parameter basis.}
\label{fig:exampleAb}
\end{figure}

Alternatively, the functional parameter can be represented by B-spline bases. Suppose that the functional parameter in model \eqref{eq:exampleA} is now expanded as a linear B-spline with a single knot at $t=0.5$, then the parameter arguments must be updated to:

\begin{verbatim}
R> pars <- c("bspline")
R> db <- c(1)
R> knotsb <- list(c(0.5))   
\end{verbatim}

and with exactly the same \texttt{pflm} call, the functions of the profile factor are updated to the functions in Figure~\ref{fig:exampleAb}.

\subsection{Functional linear regression model with an interaction} \label{subsec:example2}

An experiment is assumed to be modeled by a functional linear model that depends on two single profile factors and their interaction term. The aim is to identify a Bayesian A-optimal design of $n=12$ runs. Under these settings, the functional linear model is defined as:
\begin{equation}
\begin{split}
\label{eq:exampleB}
y_{i} & = \beta_{1} + \int_{0}^{\mathcal{T}} \beta_{2}(t) x_{i1}(t) \; dt + \int_{0}^{\mathcal{T}} \beta_{3}(t) x_{i2}(t) \; dt + \int_{0}^{\mathcal{T}} \beta_{4}(t) x_{i1}(t) x_{i2}(t) \; dt + \epsilon_{i}, \\
& \hspace{0.5cm} i = 1,2, \dots, 12, \;  t \in [0,1].
\end{split}
\end{equation}

The experimental capabilities are quadratic functions, so control of the profile factor is achieved by a quadratic B-spline, i.e., degree two B-spline, with four equally spaced interior knots. The number of runs, number of factors and specifications of the factor will be passed to \texttt{pflm} using:

\begin{verbatim}
R> nruns <- 12
R> npf <- 2
R> dx <- c(2, 2)
R> knotsx <- list(c(0.20, 0.40, 0.60, 0.80), c(0.20, 0.40, 0.60, 0.80))
\end{verbatim}

Quadratic B-spline bases are used for the parameters of the first profile factor and the interaction, and a linear B-spline basis is used for the parameter of the second profile factor. A prior specification is set via the roughness penalty approach to penalise the complexity of the functions with a smoothing parameter $\lambda=1$. The above information will be passed to \texttt{pflm} using:

\begin{verbatim}
R> pars <- c("bspline", "bspline", "bspline")
R> db <- c(2, 1, 2)
R> knotsb <- list(c(0.5), c(0.5), c(0.5))
R> lambda <- 1
\end{verbatim}

The Bayesian A-optimal design search will be for a single starting design that will be generated automatically in \texttt{pflm} by leaving \texttt{startd} in its default option. All other arguments remain at their default options. The Bayesian A-optimal design is identified via the \texttt{pflm} such that:

\begin{verbatim}
R> set.seed(1)
R> exampleB <- pflm(formula = ~ x1 + x2 + x1:x2, nsd = 1, npf = npf,
+                   tbounds = c(0, 1), nruns = nruns, dx = dx, 
+                   knotsx = knotsx, pars = pars, db = db, 
+                   knotsb = knotsb, criterion = "A", lambda = lambda)
R> print(exampleB)
\end{verbatim}

\begin{verbatim}
The number of profile factors is: 2

The number of runs is: 12

The objective criterion is: A-optimality

The objective value is: 13.33739

The number of iterations is: 12

The computing elapsed time is: 00:00:10
\end{verbatim}

\subsection{Biopharmaceutical bioreactor dynamic experiment} \label{subsec:example3}

This example is based on the Ambr250 bio-reactor experiment discussed in \citet{michaelides2021optimal}. The experiment involves four controllable factors, of which three are static factors and the fourth is a profile factor. The static factors are pH, temperature, and IVCC, and the profile factor is feed volume. The optimal design in \citet{michaelides2021optimal} aimed the estimate of linear and quadratic effects for the static factors and a quadratic function for the profile factor. However, it is in our interest here to reduce the complexity (and hence, computational burden) of the model and estimate only linear effects of the static factors and a linear function for the profile factor. At the final day of the experiment the titre content is measured. In an effort to minimise the average variance of the estimates, the application is based on a $n=12$-run A-optimal design. Thus, the functional linear model that represents the Ambr250 experiment is specified as follows: 
\begin{equation}
\label{eq:applicationequation}
\begin{split}
\text{titre}_{i} & = \beta_{1} + \int_{0}^{\mathcal{T}} \beta_{2}(t) \text{feed}_{i1}(t) \; dt +
  \int_{0}^{\mathcal{T}} \beta_{3}(t) \text{pH}(t) \; dt  + \int_{0}^{\mathcal{T}} \beta_{4}(t) \text{temp}(t) \; dt \\
& \hspace{0.2cm}    + \int_{0}^{\mathcal{T}} \beta_{5}(t) \text{IVCC}(t) \; dt  + \epsilon_{i}, \quad i = 1,2, \dots, 12, \;  t \in [0,1]
\end{split}
\end{equation}

The feed volume is dynamically controlled over time as a step function. A step function can be represented by a degree zero B-spline. Step functions are piecewise constant and represent the simplest form of non-constant functions, making them highly applicable in practice. For example, \citet{rameez:2014} utilised a step function in an Ambr bio-reactor application, implementing a temperature shift from a high to a low level when the culture reached peak viable cell density. In total, feed volume is expressed with four basis functions, resulting from the choice of three equally spaced knots. Since, pH, temperature, and IVCC are static factors, they are modeled using single constant bases functions. This is achieved via the choice of degree zero B-splines and no interior knots. 

The parameter function $\beta_{2}(t)$ that is involved with feed volume is expanded using a linear power series basis comprising three bases functions. The parameter functions $\beta_{3}(t) - \beta_{5}(t)$ that are involved with the terms of the static factors are expanded by single constant bases functions. This is achieved via the choice of degree zero power series basis and no interior knots. 

To identify an A-optimal design using \texttt{pflm}, first the settings of the model need to be specified. In context, the time bound, design runs $n$, number of factors, B-spline degree of the factors, and choice of knots are specified such that: 

\begin{verbatim}
R> tbounds <- c(0, 1)
R> nruns <- 12
R> npf <- 4
R> dx <- c(0, 0, 0, 0)
R> knotsx <- list(c(0.25, 0.50, 0.75), c(), c(), c())
\end{verbatim}

The function \texttt{pflm} gives the option of defining the starting designs through a list of designs in \texttt{startd}, or alternatively leave it to the default option in which random starts are generated automatically. In this example, the starting designs are specified. To do that, first the number of basis functions of all factors is calculated, based on the degree and number of knots above.  

\begin{verbatim}
R> nx <- rep(0, npf)
R> for (j in 1:npf) {
+    nx[j] <- dx[j] + length(knotsx[[j]]) + 1
+  }
\end{verbatim}

The starting designs are created as a list using the code lines below, with the lower and upper bounds of the factors set to their defaults ($-1$ and $1$). The number of random starts is set to \texttt{nsd}$=20$. The four factors are named $x1, x2, x3, x4$ which will then need to match the naming of the factors in the \texttt{formula} argument.

\begin{verbatim}
R> indd <- list()
R> startd <- list()
R> dlbound <- -1
R> dubound <- 1
R> nsd <- 20
R> for (c in 1:nsd) {
+    set.seed(c)
+    for (i in 1:npf) {
+      indd[[i]] <- matrix(runif(nruns * nx[i], dlbound, dubound), 
+                          nrow = nruns, ncol = nx[i])
+      names(indd)[i] <- paste0("x", i, sep="")
+    }
+    startd[[c]] <- indd
+  }
\end{verbatim}

All parameters will be expanded via power series, with the functional parameter represented by a linear power series (degree 1) and the scalar parameters represented by constant basis (degree 0). The knots selection for parameters is a list of empty vectors because no knots are specified for power series (only if parameters are expanded by B-splines), with details in the documentation of the function \texttt{pflm} and specifically arguments \texttt{pars} and \texttt{db}.

\begin{verbatim}
R> pars = c("power", "power", "power", "power")
R> db = c(1, 0, 0, 0)
R> knotsb = list(c(), c(), c(), c())
\end{verbatim}

In addition to the mandatory arguments specified above, the \texttt{formula} argument is needed to define the model's structure, the criterion is required which in that case is "A" to identify an A-optimal design, and the selection of the smoothing term \texttt{lambda} 
is set to 0 as frequentist design is needed. Optional arguments not specified are set to their defaults.

\begin{verbatim}
R> exampleC <- pflm(formula = ~ x1 + x2 + x3 + x4, nsd = nsd, npf = npf, 
+                   tbounds = tbounds, nruns = nruns, startd = startd, 
+                   dx = dx, knotsx = knotsx, pars = pars, db = db, 
+                   knotsb = knotsb, criterion = "A", lambda = 0, 
+                   dlbound = dlbound, dubound = dubound)
\end{verbatim}

\begin{figure}[ht]
\centering
\includegraphics[width=\linewidth]{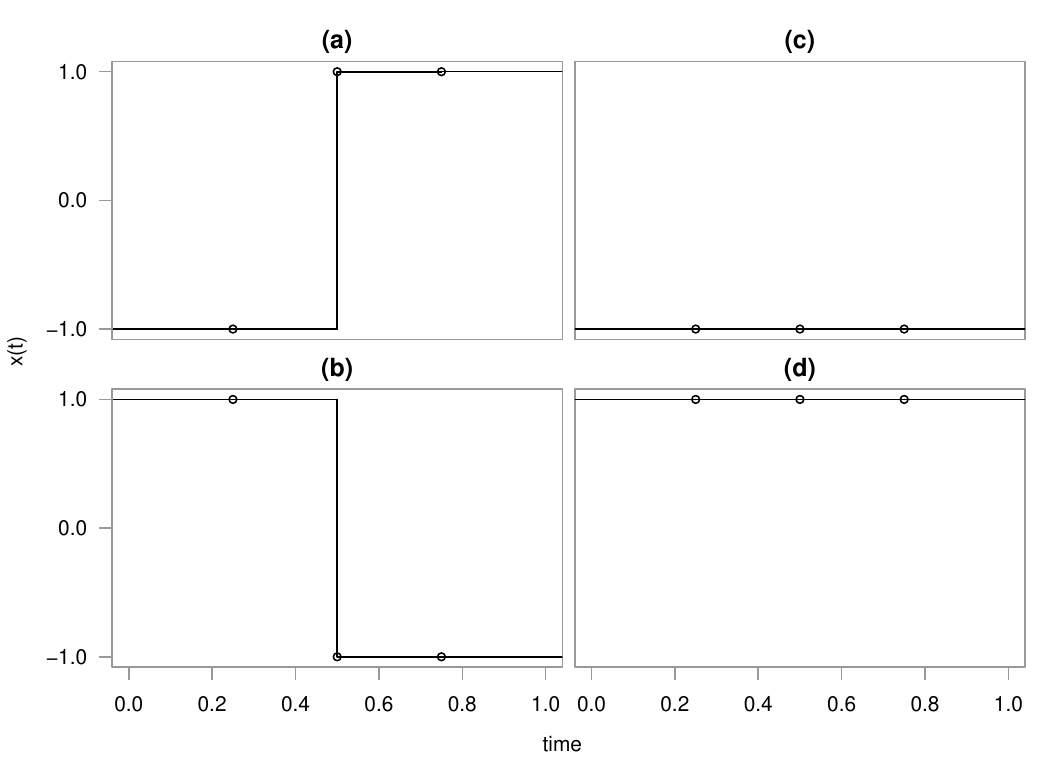}
\caption{The four unique functions of the profile factor Feed Volume in the Ambr250 bio-reactor example.}
\label{fig:exampleC}
\end{figure}

The A-optimal design consists of four unique functions of feed volume, as in Figure~\ref{fig:exampleC}. There are five runs in which the function has an increasing single step change at $t=0.5$ (from -1 to 1) and five runs in which the function has a decreasing single step change at $t=0.5$ (from 1 to -1). The remaining two runs are kept constant at the lower and upper limit. 

The optimal choice of the scalar factor values is at the boundaries with the columns of scalar factors being orthogonal. The full A-optimal design is given in Table~\ref{table:aoptimalexampleC}.

\begin{table}[ht]
\centering
\begin{tabular}{r|rrrr}
  \hline
$i$ & FV & pH & Temp. & IVCC \\
  \hline
  1  & (a) & -1.00 & 1.00 & 1.00 \\ 
  2  & (b) & -1.00 & -1.00 & 1.00 \\ 
  3  & (b) & 1.00 & -1.00 & -1.00 \\ 
  4  & (a) & 1.00 & -1.00 & 1.00 \\ 
  5  & (a) & -1.00 & 1.00 & -1.00 \\ 
  6  & (a) & -1.00 & -1.00 & -1.00 \\ 
  7  & (c) & 1.00 & 1.00 & -1.00 \\ 
  8  & (b) & -1.00 & -1.00 & -1.00 \\ 
  9  & (b) & -1.00 & 1.00 & 1.00 \\ 
  10 & (b) & 1.00 & 1.00 & 1.00 \\ 
  11 & (a) & 1.00 & -1.00 & 1.00 \\ 
  12 & (d) & 1.00 & 1.00 & -1.00 \\ 
   \hline
\end{tabular}
\caption{The 12-run A-optimal design in the Ambr250 bio-reactor example. FV is Feed Volume and Temp. is Temperature. The column of FV is labelled (a)-(d) for the functions demonstrated in Figure~\ref{fig:exampleC}.}
\label{table:aoptimalexampleC}
\end{table}

\subsection{Functional logistic regression model} \label{subsec:example4}

An experiment is assumed to be modeled by a functional logistic model that depends on a single profile factor. The aim is to identify a pseudo-Bayesian D-optimal design of $n=12$ runs. Thus, the functional generalised linear model of interest is defined as:
\begin{flalign}
\label{eq:logistic}
& y_{i} \sim \text{Bernoulli}(\kappa_{i})  \nonumber\\
& \eta_{i} = \beta_{1} +  \int_{0}^{T} \beta_{2}(t) x_{i1}(t) \; dt, \quad i=1,2, \dots, 12, \;  t \in [0,1],
\end{flalign}
with link function $g(\kappa_{i})= \log \big( \kappa_{i} / (1 - \kappa_{i}) \big)$. 

Given a functional logistic model, the \texttt{family} in the \texttt{pfglm} function must be chosen to be:

\begin{verbatim}
R> family_binomial <- binomial
\end{verbatim}

The experimental capabilities are step functions that can change up to three times, so control of the profile factor is represented by a degree zero B-spline with three equally spaced interior knots. The number of runs, number of factors and specifications of the factor will be passed to \texttt{pfglm} using:

\begin{verbatim}
R> nruns <- 12
R> npf <- 1
R> dx <- c(0)
R> knotsx <- list(c(0.25, 0.50, 0.75))                				
\end{verbatim}

A linear power basis is assigned to the functional parameter that corresponds to the single profile factor. The smoothing parameter $\lambda=1$ is irrelevant for a linear basis because the second derivatives of linear terms are zero, and thus, the roughness matrix has all values being zero. With that being said, \texttt{lambda} can be at its default choice, which is \texttt{lambda}$=0$. The above information will be passed to \texttt{pfglm} using:

\begin{verbatim}
R> pars <- c("power")
R> db <- c(1)
R> knotsb <- list(c())   
\end{verbatim}

As discussed in Section \ref{subsec:pseudo}, optimal experimental designs for functional generalised linear models require prior information about the model parameters. The logistic model in \eqref{eq:logistic} has two unknown parameters, $\beta_{1}$ and $\beta_{2}(t)$. The former is the intercept, then, a single constant basis function is needed. Using basis expansions from \eqref{eq:basesbnu}, $\beta_{2}$ is expanded via linear basis, and thus, the parameter vector from \eqref{eq:extendedmodelglm} is,
\begin{equation}
\label{eq:singleprofilelinearthetas}
\bm{\theta}^{T}=
\begin{pmatrix}
\theta_{1} & \bm{\theta}_{2}
\end{pmatrix} = 
\begin{pmatrix}
\theta_{1} &  \theta_{21} & \theta_{22},
\end{pmatrix},
\end{equation}
for which independent uniform prior distributions are assumed such that:
\begin{equation}
\label{eq:priorthetas01linear}
\theta_{1} \sim U(-2,2), \quad \bm{\theta}_{2} \sim U(3,9).
\end{equation}
Additionally, a pseudo-Bayesian D-optimal design is identified using the quadrature approximation. The approximation method and Uniform prior specification of the parameters will be passed to \texttt{pfglm} using: 

\begin{verbatim}
R> approx_method <- c("quadrature")
R> unif_lower <- c(-2, 3)
R> unif_upper <- c(2, 9)
R> unif_parms <- matrix(c(c(unif_lower), c(unif_upper)), nrow = 2, byrow = TRUE)
R> prior_list <- list(unifbound = unif_parms)              				
\end{verbatim}

since for quadrature method of approximation the \texttt{prior} argument in \texttt{pfglm} must be a list of single component named "unifbound" which is, a vector if all parameters have the same bounds, or a 2-row matrix if parameters have different bounds. 

The pseudo-Bayesian D-optimal design search will be for a single starting design that will be generated automatically in \texttt{pfglm} by leaving \texttt{startd} in its default option. All other arguments remain at their default options. The pseudo-Bayesian D-optimal design is identified via \texttt{pfglm} such that:

\begin{verbatim}
R> set.seed(2)
R> exampleD <- pfglm(formula = ~ 1 + x1, nsd = 1, npf = npf, 
+                tbounds = c(0, 1), nruns = nruns, dx = dx, 
+                knotsx = knotsx, pars = pars, db = db, 
+                knotsb = knotsb, criterion = "D", 
+                family = family_binomial, method = approx_method, 
+                prior = prior_list)
R> print(exampleD)
\end{verbatim}

\begin{figure}[ht]
\centering
\includegraphics[width=\linewidth]{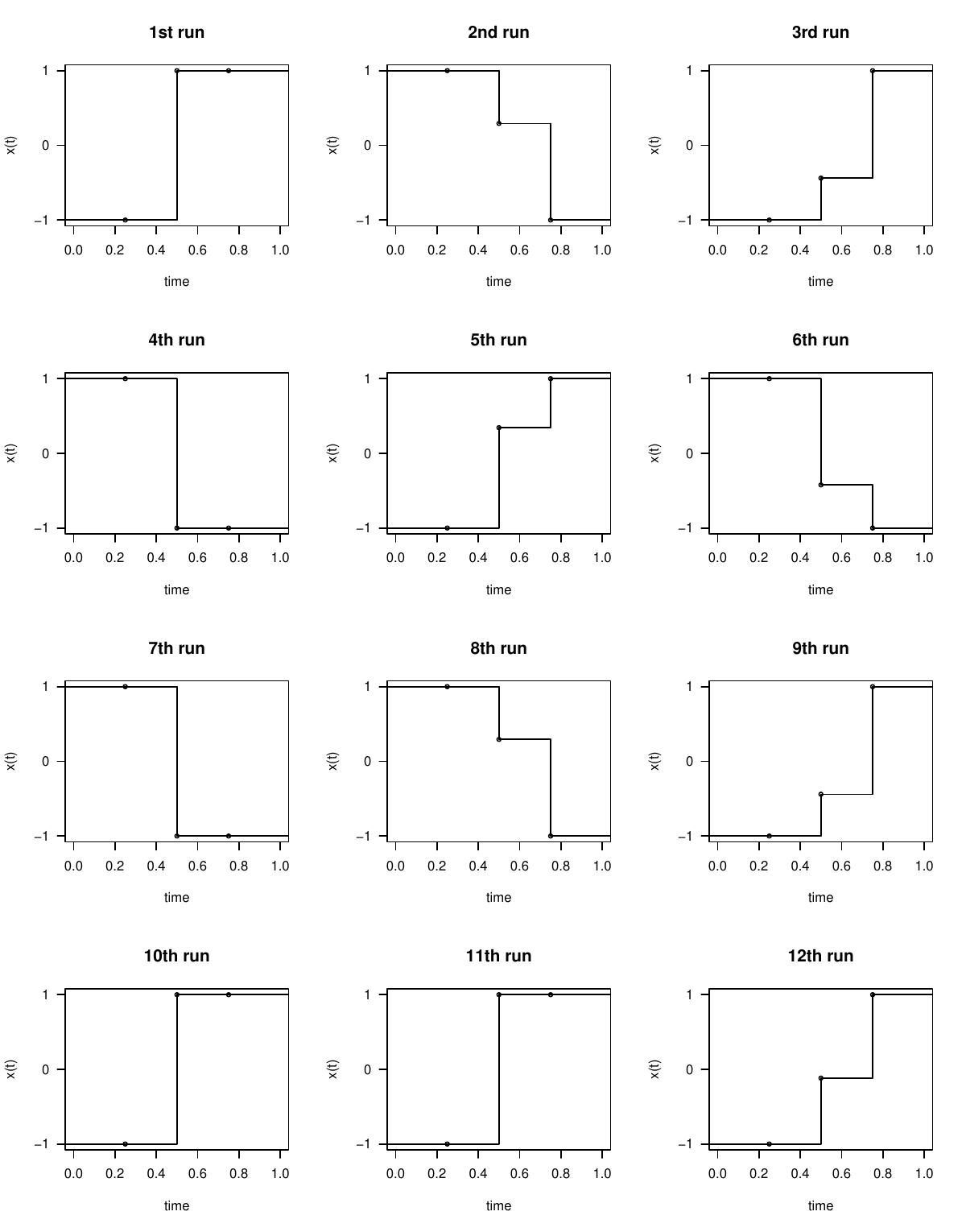}
\caption{Pseudo-Bayesian 12-run D-optimal design for the functional logistic model.}
\label{fig:exampleD}
\end{figure}

\begin{verbatim}
The number of profile factors is: 1

The number of runs is: 12

The objective criterion is: D-optimality

The objective value is: 847.976

The number of iterations is: 31

The method of approximation is: quadrature

The family distribution and the link function are: binomial and logit

The computing elapsed time is: 00:00:01
\end{verbatim}

and the functions of the profile factor resulting from the pseudo-Bayesian 12-run D-optimal design identified are plotted via:

\begin{verbatim}
R> plot(exampleD)
\end{verbatim}

and then input the number of the profile factor to plot (in this case 1) to the question appearing "Which profile factor to plot?". The functions of the profile factor as shown in Figure~\ref{fig:exampleD}.

\subsection{Functional Poisson regression model} \label{subsec:example5}

An experiment is assumed to be modeled by a functional Poisson model that depends on two profile factors. The aim is to identify a pseudo-Bayesian A-optimal design of $n=12$ runs. Thus, the functional generalised linear model of interest is defined as:
\begin{flalign}
\label{eq:poisson}
& y_{i} \sim \text{Poisson}(\kappa_{i})  \nonumber\\
& \eta_{i} = \beta_{1} + \int_{0}^{\mathcal{T}}  x_{i1}(t) \beta_{2} (t) \; dt + \int_{0}^{\mathcal{T}}  x_{i2}(t) \beta_{3} (t) \; dt, \quad i=1,2, \dots, 12, \;  t \in [0,1],
\end{flalign}
with link function $g(\kappa_{i})= \log( \kappa_{i})$. 

Given a functional Poisson model, the \texttt{family} in the \texttt{pfglm} function must be chosen to be:

\begin{verbatim}
R> family_Poisson <- poisson
\end{verbatim}

It is assumed that the two profile factors have different capabilities which must be reflected in their settings. The first profile factor $x_{i1}(t)$ can be flexible cubic functions, and so control of $x_{i1}(t)$ is represented by a cubic B-spline, i.e., degree three B-spline, with four equally spaced interior knots. The second profile factor $x_{i2}(t)$ can be step functions with a single step change at $t=0.5$, and so control of $x_{i2}(t)$ is represented by a degree zero B-spline with a single interior knot at $t=0.5$. The number of runs, number of factors and the specifications of the two profile factors will be passed to \texttt{pfglm} using:

\begin{verbatim}
R> nruns <- 12
R> npf <- 2
R> dx <- c(3, 0)
R> knotsx <- list(c(0.20, 0.40, 0.60, 0.80), c(0.5))             				
\end{verbatim}

Quadratic and linear power bases are assigned to the functional parameters $\beta_{2}(t)$ and $\beta_{3}(t)$, respectively. The smoothing parameter choice is $\lambda=1$ that will affect the quadratic parameters, but not the linear parameters. The above information will be passed to \texttt{pfglm} using:

\begin{verbatim}
R> pars <- c("power", "power")
R> db <- c(2, 1)
R> knotsb <- list(c(), c())   
R> lambda <- 1
\end{verbatim}

The pseudo-Bayesian A-optimal design in this example is identified using the Monte Carlo approximation with a normal prior of mean 0 and variance 2. The approximation method and the prior specification will be passed to \texttt{pfglm} using: 

\begin{verbatim}
R> approx_method <- c("MC")
R> prmc <- function(B, Q){
+    matrix(rnorm(B * Q, mean = 0, sd = sqrt(2)), nrow = B, ncol = Q)
+  }             				
\end{verbatim}

since for MC method of approximation the \texttt{prior} argument in \texttt{pfglm} must be a function of two arguments, B and Q, both of which are integers. Here, B corresponds to \texttt{pfglm}'s argument \texttt{B} that specifies the size of the Monte Carlo samples. Argument Q denotes the total number of bases functions for the functional parameters that is calculated automatically within \texttt{pfglm} based on the settings of the model. The prior function is required to generate a matrix of dimensions B by Q, containing random samples drawn from the prior distribution of the parameters. The size of the Monte Carlo samples is set to the default option which is \texttt{B}$=10000$ when the approximation method is MC. 

The function \texttt{pfglm}, as in \texttt{pflm}, gives the option of defining the starting designs through a list of designs in \texttt{startd}, or alternatively leave it to the default option in which random starts are generated automatically, as shown in the logistic model example before. In this example, the starting design is specified. To do that, first the number of basis functions of all factors are calculated, based on the their degree and number of knots.  

\begin{verbatim}
R> nx <- rep(0, npf)
R> for (j in 1:npf) {
+    nx[j] <- dx[j] + length(knotsx[[j]]) + 1
+  }
\end{verbatim}

A single random start is assumed, i.e., \texttt{nsd}$=1$. The two profile factors are named $x1, x2$ which need to match the naming of the factors in the \texttt{formula} argument. 

\begin{verbatim}
R> indd <- list()
R> startd <- list()
R> dlbound <- -1
R> dubound <- 1
R> nsd <- 1
R> set.seed(150)
R> for (c in 1:nsd) {
+    for (i in 1:npf) {
+      indd[[i]] <- matrix(runif(nruns * nx[i], dlbound, dubound), 
+                          nrow = nruns, ncol = nx[i])
+      names(indd)[i] <- paste0("x", i, sep="")
+    }
+    startd[[c]] <- indd
+  }
\end{verbatim}

All other arguments remain at their default options. The pseudo-Bayesian A-optimal design is identified via \texttt{pfglm} such that:

\begin{verbatim}
R> exampleE <- pfglm(formula = ~ 1 + x1 + x2, nsd = nsd, npf = npf, 
+                  startd=startd, tbounds = c(0, 1), nruns = nruns, 
+                  dx = dx, knotsx = knotsx, pars = pars, db = db, 
+                  knotsb = knotsb, criterion = "A", prior = prmc,  
+                  family = family_Poisson, method = approx_method, 
+                  dlbound = dlbound, dubound = dubound)
R> print(exampleE)
\end{verbatim}

\begin{verbatim}
The number of profile factors is: 2

The number of runs is: 12

The objective criterion is: A-optimality

The objective value is: 2.212573

The number of iterations is: 4

The method of approximation is: MC

The family distribution and the link function are: poisson and log

The computing elapsed time is: 00:00:34
\end{verbatim}

\begin{figure}[ht]
\centering
\includegraphics[width=\linewidth]{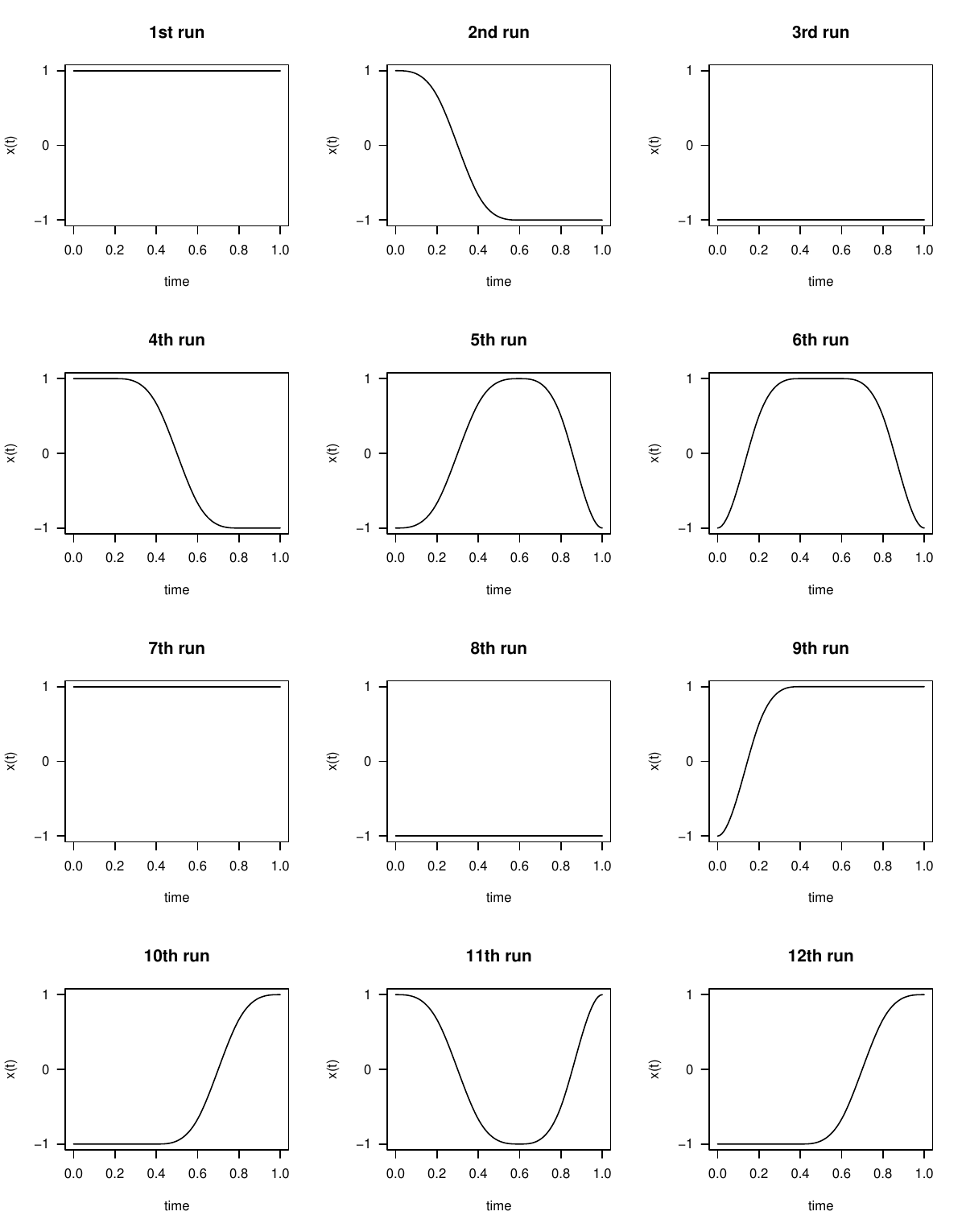}
\caption{Pseudo-Bayesian 12-run A-optimal design for the functional Poisson model and profile factor $x_{i1}(t)$.}
\label{fig:exampleE1}
\end{figure}

\begin{figure}[ht]
\centering
\includegraphics[width=\linewidth]{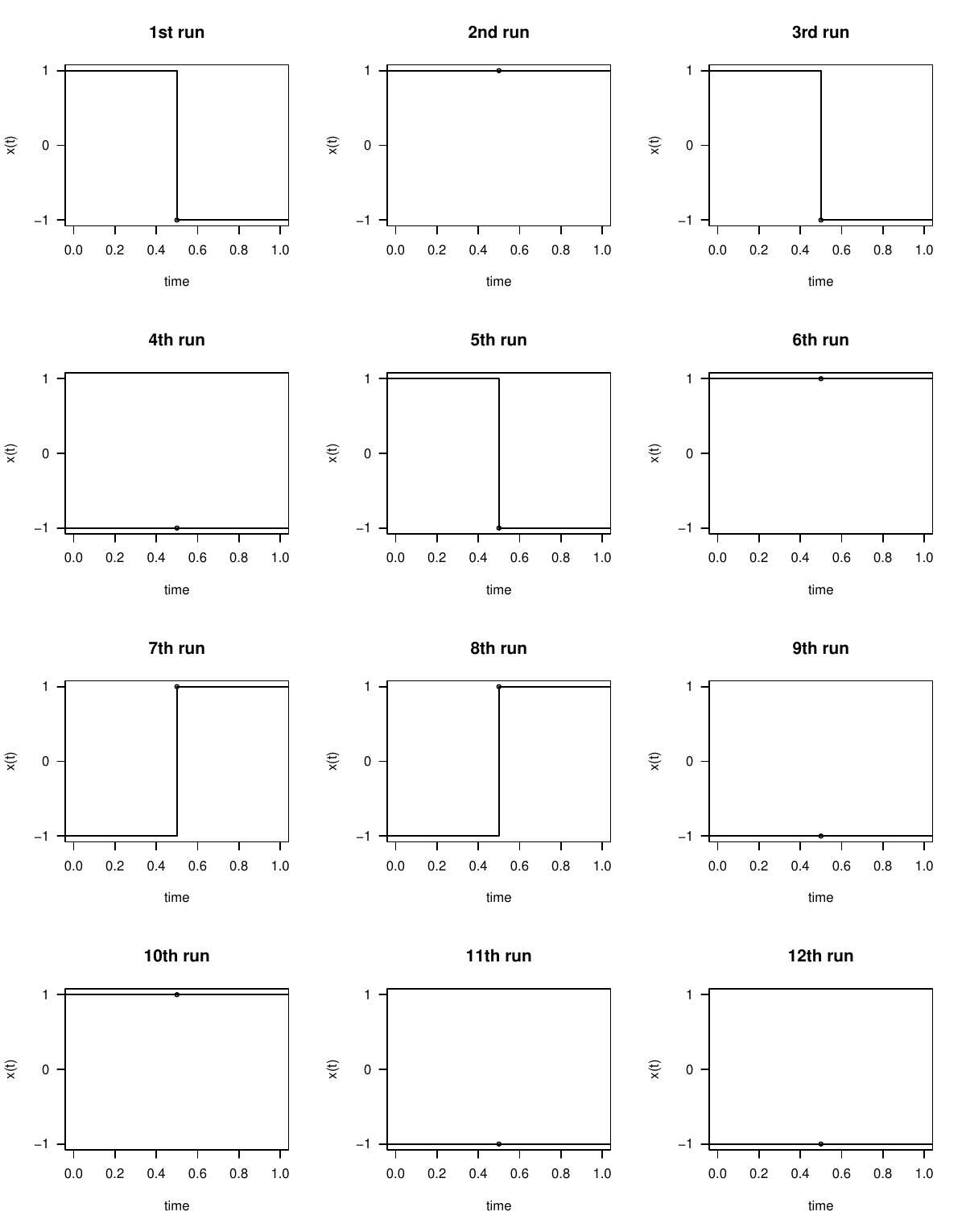}
\caption{Pseudo-Bayesian 12-run A-optimal design for the functional Poisson model and profile factor $x_{i2}(t)$.}
\label{fig:exampleE2}
\end{figure}

The functions of the profile factors resulting from the pseudo-Bayesian 12-run A-optimal design identified can be seen via plotting the \texttt{fglm} object generated such that:

\begin{verbatim}
R> plot(exampleE)
\end{verbatim}

and then input the number of the profile factor to plot to the question appearing "Which profile factor to plot?". If the input is 1, the functions of the profile factor $x_{i1}(t)$ are plotted as shown in Figure~\ref{fig:exampleE1}, and if the input is 2, the functions of the profile factor $x_{i2}(t)$ are plotted as shown in Figure~\ref{fig:exampleE2}.

\section{Concluding remarks} \label{sec:conclusion}

The contribution of the development of the \texttt{R} package \texttt{fdesigns} is its capability to generate Bayesian optimal experimental designs for models where factors and parameters are functions of time, specifically for functional linear and generalised linear models. The methodology by \citet{michaelides2021optimal} using bases functions is implemented in \texttt{fdesigns} allowing the restriction of the function space, thus, simplifying parameter estimation and enhancing the efficiency of the design.

The methodology for functional linear and generalised linear models is implemented by the functions \texttt{pflm} and \texttt{pfglm}, respectively. The latter functions utilise the coordinate exchange algorithm and approximation methods (for generalised models), enabling users to find optimal designs tailored to their specific needs. The parallel computing feature enhances the utility of \texttt{fdesigns}'s functions by allowing multiple starting designs, thus increasing the likelihood of finding better locally, or even globally, Bayesian optimal designs. The integration of \texttt{C++} through the packages \texttt{Rcpp} and \texttt{RcppArmadillo} ensures that the package \texttt{fdesigns} performs efficiently and at low computational burden, even for computationally intensive tasks. 

Practical applications of \texttt{fdesigns} are provided through five examples, showcasing the different options, including interactions, polynomial effects, roughness penalty, etc, as well as printing and plotting features, across various scenarios. Although the examples presented in the paper are deliberately kept relatively simple to provide computationally fast demonstration of the functions, \texttt{fdesigns} can be used to identify optimal designs for more complex scenarios with multiple factors or higher order B-splines.  

Future plans are to extend the package to accommodate more types of models and additional design criteria, including criteria tailored to profile factors, to broaden \texttt{fdesigns}'s applicability and impact in the area of optimal experimental designs.

\bibliographystyle{apalike} 

\bibliography{refs}


\newpage

\begin{appendix}

\section{Tables with descriptions of the output components of the functions} \label{app:tables}

\begin{longtable}{p{2.5cm}p{12cm}}
\hline
 \textbf{Output} & \textbf{Description}  \\
 \hline
\texttt{objval} & The objective value of the final design found from \texttt{pflm}. \\
\texttt{design} & The final design found by \texttt{pflm}. The final design is a list of length equal to the number of profile factors, exactly as the starting design \texttt{startd}.\\
\texttt{nits} & The total number of iterations needed to identify the final design.\\
\texttt{time} & The computational elapsed time in finding the final design.\\
\texttt{startd} & If starting designs were passed as an argument in \texttt{pflm}, then this is the starting design from the argument \texttt{startd} that led to the final design. If no starting designs were passed to \texttt{pflm}, this is the starting design generated randomly by \texttt{pflm} that led to the final design.\\
\texttt{tbounds} & The argument \texttt{tbounds} .\\
\texttt{npf} & The argument \texttt{npf}.\\
\texttt{criterion} & The argument \texttt{criterion}.\\
\texttt{nruns} & The argument \texttt{nruns}.\\
\texttt{formula} & The argument \texttt{formula}.\\
\texttt{dx} & The argument \texttt{dx}.\\
\texttt{knotsx} & The argument \texttt{knotsx}.\\
\texttt{lambda} & The argument \texttt{lambda}.\\
\texttt{dbounds} & A vector of length 2, containing the arguments \texttt{dlbound} and \texttt{dubound}.\\
\texttt{bestrep} & A scalar value indicating the repetition that led to the final design.\\
\texttt{allobjvals} & A vector of length equal to \texttt{nsd}, representing the objective value from all of the repetitions.\\
\texttt{alldesigns} & A list of length equal to \texttt{nsd} of all the final designs. Each component of the list is a list of length equal to \texttt{npf} representing the final design in each repetition of the coordinate exchange algorithm.\\
\texttt{allstartd} & If starting designs were passed as an argument in \texttt{pflm}, then this is the argument. If no starting designs were passed to \texttt{pflm}, this is the starting designs generated randomly by \texttt{pflm}.\\
 \hline 
\caption{The output components of \texttt{pflm} with a description for each.}
\label{table:outputspflm}
\end{longtable}

\begin{longtable}{p{3cm}p{11.5cm}}
\hline
 \textbf{Output} & \textbf{Description}  \\
 \hline
\texttt{objective.value} & {The objective value of the final design found from \texttt{pfglm}.} \\
\texttt{design} & {The final design found from \texttt{pfglm}. The final design is a list of length equal to the number of profile factors, 
        exactly as the starting design \texttt{startd}.} \\
\texttt{n.iterations} & {The total number of iterations needed to identify the final design.} \\
\texttt{time} & {The computational elapsed time in finding the final design.} \\
\texttt{startd} & {If starting designs were passed as an argument in \texttt{pfglm}, then this is the argument \texttt{startd}.
        If no starting designs were passed to \texttt{pfglm}, this is the starting design generated randomly by \texttt{pfglm}.} \\
\texttt{tbounds} & {The argument \texttt{tbounds}.} \\
\texttt{npf} & {The argument \texttt{npf}.} \\
\texttt{criterion} & {The argument \texttt{criterion}.} \\
\texttt{nruns} & {The argument \texttt{nruns}.} \\
\texttt{formula} & {The argument \texttt{formula}.} \\
\texttt{family} & {A vector of length equal to 2, containing the family and the link function.} \\
\texttt{method} & {The argument \texttt{method}.} \\
\texttt{B} & {The argument \texttt{B}.} \\
\texttt{prior} & {The argument \texttt{prior}.} \\
\texttt{dx} & {The argument \texttt{dx}.} \\
\texttt{knotsx} & {The argument \texttt{knotsx}.} \\
\texttt{lambda} & {The argument \texttt{lambda}.} \\
\texttt{dbounds} & {A vector of length 2, containing the arguments \texttt{dlbound} and \texttt{dubound}.} \\
\texttt{bestrep} & {A scalar value indicating the repetition that led to the final design.} \\
\texttt{allobjvals} & {A vector of length equal to nsd, representing the objective value from all of the repetitions.} \\
\texttt{alldesigns} & {A list of length equal to \texttt{nsd} of all the final designs. Each component of the list is a list of length equal to \texttt{npf}              representing the final design in each repetition of the coordinate exchange algorithm.} \\
\texttt{allstartd} & {If starting designs were passed as an argument in \texttt{pfglm}, then this is the argument. If no starting designs were passed to                     \texttt{pfglm}, this is the starting designs generated randomly by \texttt{pfglm}.} \\
 \hline
\caption{The output components of \texttt{pfglm} with a description for each.}
\label{table:outputspfglm}
\end{longtable}

\end{appendix}


\end{document}